\documentclass[showpacs,amsmath,amssymb,prb,superscriptaddress,twocolumn]{revtex4}
\usepackage[T1]{fontenc}
\usepackage{amssymb,amsmath}
\usepackage{times}
\usepackage{graphicx}
\usepackage{wasysym}
\usepackage{subfigure}
\usepackage[sort&compress]{natbib}
\usepackage{bm}
\usepackage{overpic}
\usepackage{color}
\usepackage{multirow}

\begin{document}

\title{Repulsive Casimir forces with finite-thickness slabs}

\author{R.~Zhao}
\affiliation{Ames Laboratory and Department of Physics and Astronomy,
             Iowa State University, Ames, Iowa 50011, USA}
\affiliation{Applied Optics Beijing Area Major Laboratory, Department of Physics,
Beijing Normal University, Beijing 100875, China}

\author{Th.~Koschny}
\affiliation{Ames Laboratory and Department of Physics and Astronomy,
             Iowa State University, Ames, Iowa 50011, USA}
\affiliation{Institute of Electronic Structure and Laser, FORTH,
             Department of Materials Science and Technology, University of Crete, Heraklion, 71110 Crete, Greece}

\author{E.~N.~Economou}
\affiliation{Institute of Electronic Structure and Laser, FORTH,
             Department of Materials Science and Technology, University of Crete, Heraklion, 71110 Crete, Greece}

\author{C.~M.~Soukoulis}
\affiliation{Ames Laboratory and Department of Physics and Astronomy,
             Iowa State University, Ames, Iowa 50011, USA}
\affiliation{Institute of Electronic Structure and Laser, FORTH,
             Department of Materials Science and Technology, University of Crete, Heraklion, 71110 Crete, Greece}
\date{\today}


\begin{abstract}
We use the extended Lifshitz theory to study the behaviors of the Casimir forces between finite-thickness effective medium slabs. We first study the interaction between a semi-infinite Drude metal and a finite-thickness magnetic slab with or without substrate. For no substrate, the large distance $d$ dependence of the force is repulsive and goes as $1/d^5$; for the Drude metal substrate, a stable equilibrium point appears at an intermediate distance which can be tuned by the thickness of the slab. We then study the interaction between two identical chiral metamaterial slabs with and without substrate. For no substrate, the finite thickness of the slabs $D$ does not influence significantly the repulsive character of the force at short distances, while the attractive character at large distances becomes weaker and behaves as $1/d^6$; for the Drude metal substrate, the finite thickness of the slabs $D$ does not influence the repulsive force too much at short distances until $D=0.05\lambda_0$.
\end{abstract}




\maketitle
\section{Introduction}
Arising from the quantum fluctuations of the vacuum field, when two neutral parallel conducting surfaces separated by the vacuum are very close to each other, they generate an attractive force between them given by $F=-\frac{\hbar c \pi^2 A}{240 d^4}$ and called \textit{Casimir force} after the founder Casimir.\cite{Casimir} The Casimir force becomes more pronounced if the dimension goes to nanoscale. It will lead to stiction and adhesion on the surface,\cite{Buks, Serry} which is a challenge for flexibly operating the Micro/Nanoelectromechanical system devices. Later, especially recently, people were/are pursuing different methods to control the Casimir force so as to obtain a repulsive force: immersing two objects characterized by the dielectric permittivities $\epsilon_1(i\xi)$ and  $\epsilon_2(i\xi)$ in a fluid with $\epsilon_3(i\xi)$ (satisfying $\epsilon_1(i\xi)<\epsilon_3(i\xi)<\epsilon_2(i\xi)$),\cite{Dzyaloshinskii, Munday} using a special geometry,\cite{Rodriguez} an electric ($\epsilon>\mu$) plate together with a magnetic ($\mu>\epsilon$) plate,\cite{Boyer, Kenneth} two interacting plates sandwiching a perfect lens,\cite{Leonhardt} or resorting to strong chirality materials.\cite{ZhaoPRL, ZhaoPRB} Only for the first two proposals, natural materials can be utilized, while for the others, they all need some exotic materials, i.e., strong magneto-dielectric response materials,\cite{Rosa1, Rosa2, Yannopapas} perfect lens,\cite{Leonhardt} strong chiral metamaterials.\cite{ZhaoPRL, ZhaoPRB} These materials do not exist in nature and can only potentially be made artificially. This type of material is called \textit{metamaterial}.\cite{metamaterial1} Under current technologies, the thickness of these metamaterials can not be made very large especially at the optical regime.\cite{metamaterial2, metamaterial3} The thickest optical negative index metamaterial so far is only around half of the operating wavelength.\cite{Valentine} What we can obtain is just a finite-thickness artificial metamaterial slab with or without a substrate. Therefore, in this paper, we study the behaviors of the repulsive Casimir forces with finite-thickness effective medium slabs for two of the aforementioned proposals: with strong magneto-dielectric response materials \cite{Rosa1, Rosa2, Yannopapas} and with strong chiral metamaterials.\cite{ZhaoPRL, ZhaoPRB}

First we briefly introduce the extended Lifshitz theory which is valid for chiral metamaterials as well. Lifshitz \cite{Lifshitz} generalized the calculation of Casimir force between two semi-infinite planar and parallel objects $1$ and $2$ characterized by frequency-dependent
dielectric functions $\epsilon_1(\omega)$ and $\epsilon_2(\omega)$.
Later there was further extension to general
bi-anisotropic media.\cite{Parsegian,Barash,Philbin} The formula for the force or
the interaction energy per unit area can be expressed in terms of
the reflection amplitudes $r_j^{ab}$ ($j=1,2$),\cite{Lambrecht} at the interface between vacuum and the object $j$, giving the ratio of the reflected EM wave of
polarization \textit{a} by the incoming wave of polarization
\textit{b}. Each \textit{a} and \textit{b} stands for either
electric (TM or $p$) or magnetic (TE or $s$) waves. The frequency
integration is performed along the imaginary axis by setting
$\omega=i\xi$. The interaction energy per unit
area is given by
\begin{equation}\label{Casimir energy}
\frac{E(d)}{A}=\frac{\hbar}{2\pi}\int_0^{+\infty} d\xi\int\frac{d^2\mathbf{k}_\parallel}{(2\pi)^2}\ln\det\mathbf{G},
\end{equation}
where $\mathbf{G}=1-\mathbf{R}_1\cdot\mathbf{R}_2e^{-2K_0d}$,
\begin{equation}\label{reflection elements}
\mathbf{R}_j=\left\lvert\begin{array}{cc}
r_j^\textrm{ss}&r_j^\textrm{sp}\\r_j^\textrm{ps}&r_j^\textrm{pp}
\end{array}\right\rvert,
\end{equation}
and $K_0=\sqrt{\mathbf{k}_\parallel^2+\epsilon_0\mu_0\xi^2}$; $\epsilon_0$ and $\mu_0$ are the permittivity and permeability of free space, and $d$ is the distance between the two interacting plates. A negative/positive slope of $E(d)$ corresponds to a repulsive/attractive force.

For a finite-thickness isotropic achiral slab $j$ with a semi-infinite isotropic achiral substrate medium $j'$, the reflection elements are the results of the multi-scattering by the finite slab and written as
\begin{equation}\label{reflection elements finite isotropic slab}
r_j^{ab}=\frac{r_{0j}^{ab}+r_{jj'}^{ab}e^{-2K_j d_j}}{1+r_{0j}^{ab}r_{jj'}^{ab}e^{-2K_j d_j}},
\end{equation}
where $d_j$ is the thickness of the slab $j$, $K_j=\sqrt{\mathbf{k}_\parallel^2+\epsilon_0\mu_0\epsilon_{rj}\mu_{rj}\xi^2}$ , and $\epsilon_{rj}$ and $\mu_{rj}$ are the relative permittivity and permeability of the medium $j$. In $r_{mn}^{ab}$, the superscripts $a$ and $b$ are defined the same way as in Eq. (\ref{reflection elements}) and the subscripts $m$ and $n$ denote that the light is incident from the medium $m$ to $n$. $0$ means vacuum.
$r_{mn}^{ab}$ are given as \cite{Born}
\begin{subequations}\label{reflection elements infinite isotropic}
\begin{align}
r_{mn}^{ss}&=(\mu_{rn} K_m-\mu_{rm} K_n)/(\mu_{rn} K_m+\mu_{rm} K_n),\\
r_{mn}^{pp}&=(\epsilon_{rn} K_m-\epsilon_{rm} K_n)/(\epsilon_{rn} K_m+\epsilon_{rm} K_n),\\
r_{mn}^{sp}&=r_{mn}^{ps}=0.
\end{align}
\end{subequations}

For a finite-thickness isotropic chiral slab $j$ with a semi-infinite isotropic achiral substrate medium $j'$, the nondiagonal terms, $r^{sp}$ and $r^{ps}$, are nonzero. The total reflection matrix can be written as\cite{Cory}
\begin{equation}\label{reflection elements finite chiral slab}
\mathbf{R}_j=\mathbf{R}_{0j}+\mathbf{T}_{j0}\mathbf{\Delta}_j \mathbf{R}_{jj'} \mathbf{\Delta}_j [\mathbf{I}-\mathbf{R}_{j0}\mathbf{\Delta}_j \mathbf{R}_{jj'} \mathbf{\Delta}_j]^{-1} \mathbf{T}_{0j},
\end{equation}
where $\mathbf{I}$ is the unit matrix and
\begin{equation}\label{delta}
\mathbf{\Delta}_j=\left\lvert\begin{array}{cc}
e^{-K_{j+}d_j}&0\\0&e^{-K_{j-}d_j}
\end{array}\right\rvert,
\end{equation}
where $K_{j\pm}=\sqrt{\mathbf{k}_\parallel^2+n_{j\pm}^2(i\xi)\xi^2/c^2}$ and
$n_{j\pm}(i\xi)=\sqrt{\epsilon_{rj}(i\xi)\mu_{rj}(i\xi)}\pm$$\kappa_j(i\xi)$.
$\epsilon_{rj}(i\xi)$ and
$\mu_{rj}(i\xi)$ are the relative permittivity and permeability of the chiral slab $j$, respectively, and $\kappa_j(i\xi)$ is the chirality coefficient; $c$ is the velocity of the light in vacuum. The matrices $\mathbf{R}_{mn}$ and $\mathbf{T}_{mn}$ are the reflection and transmission matrices at the interface of the medium $m$ and $n$. The subscripts $m$ and $n$ still denote that the incident light is from the medium $m$ to $n$. The detailed expressions of these matrices' elements can be found in Ref. [26].

\section{Repulsive Casimir Forces with Magnetic Slabs}
\begin{figure*}[htb!]
\begin{overpic}[width=0.45\textwidth]{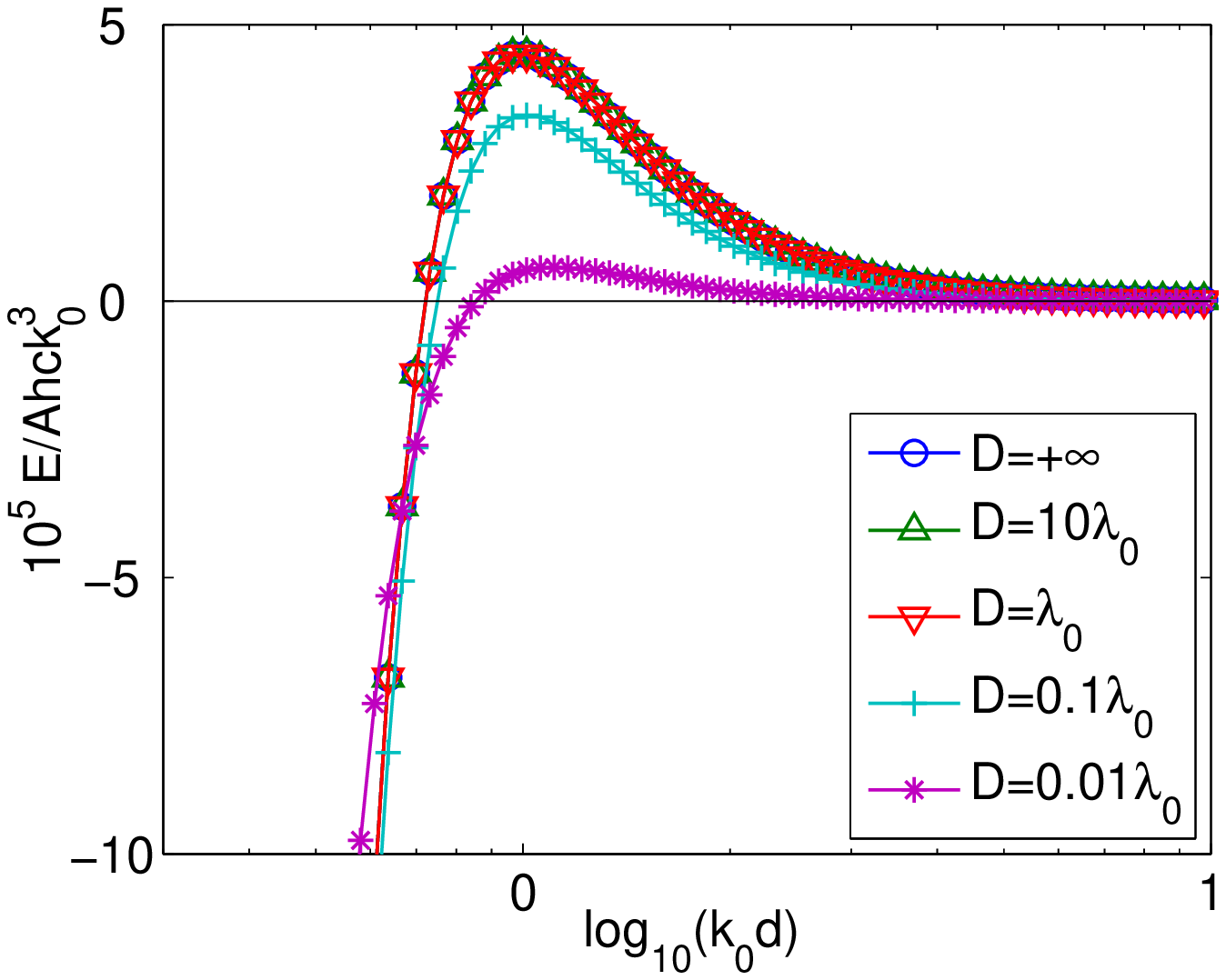}
\put(16,15){(a)}
\end{overpic}
\hspace{6mm}
\begin{overpic}[width=0.45\textwidth]{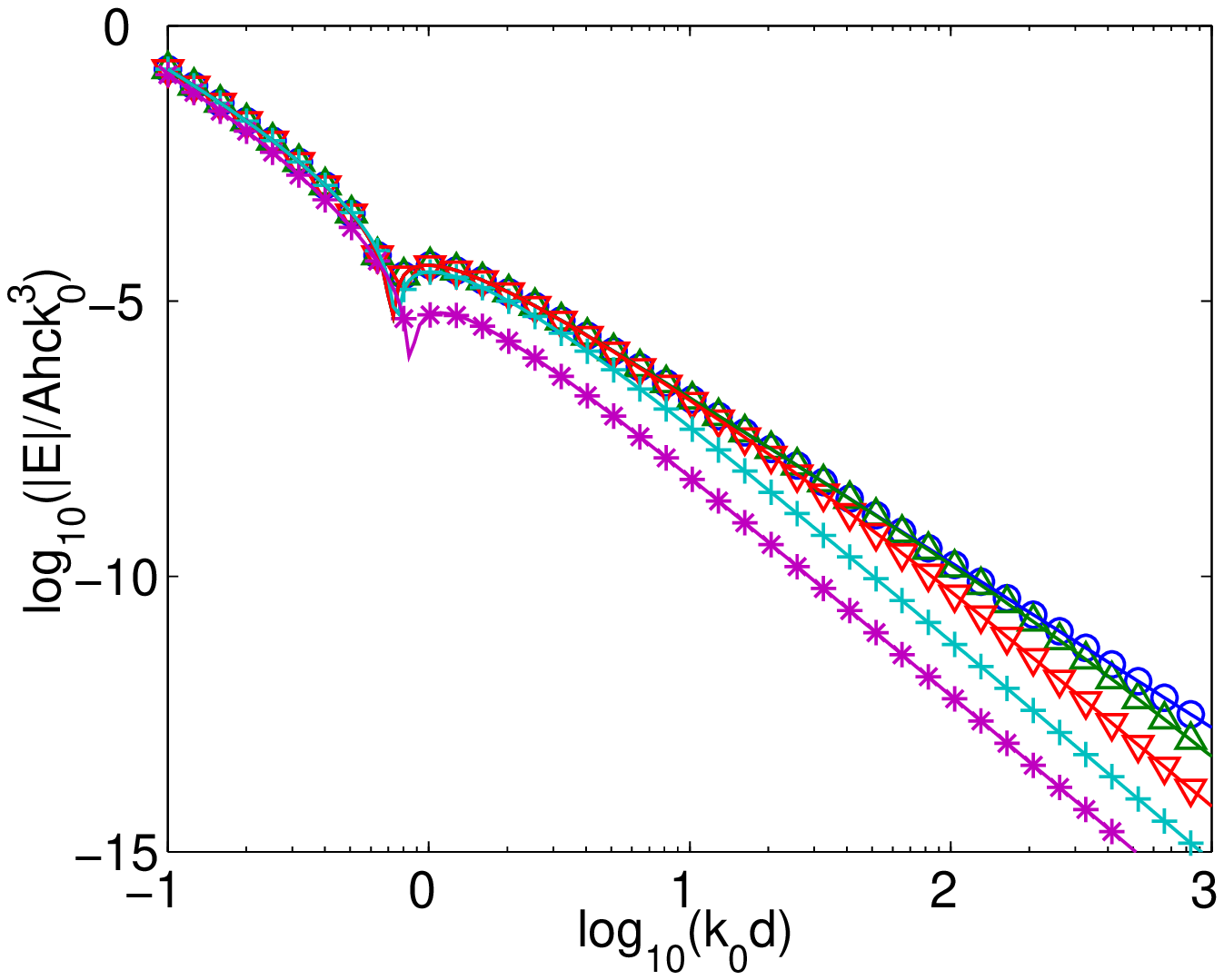}
\put(54,58){\includegraphics[width=0.19\textwidth]{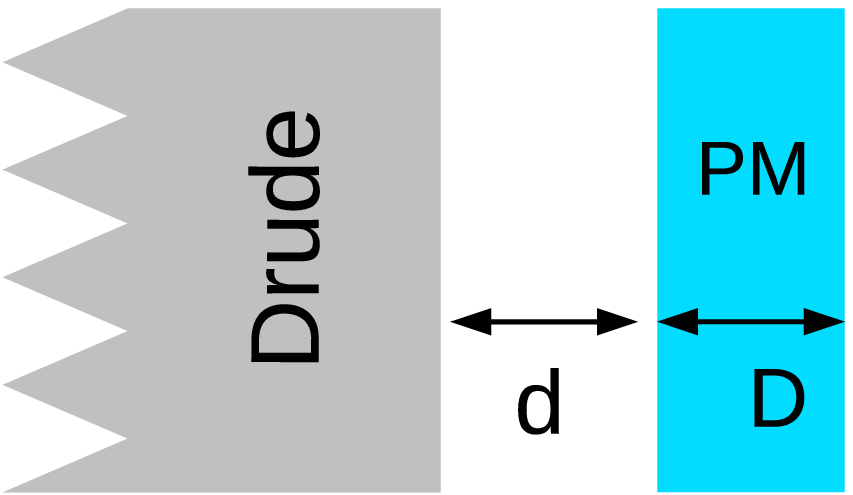}}
\put(16,15){(b)}
\end{overpic}

\begin{overpic}[width=0.45\textwidth]{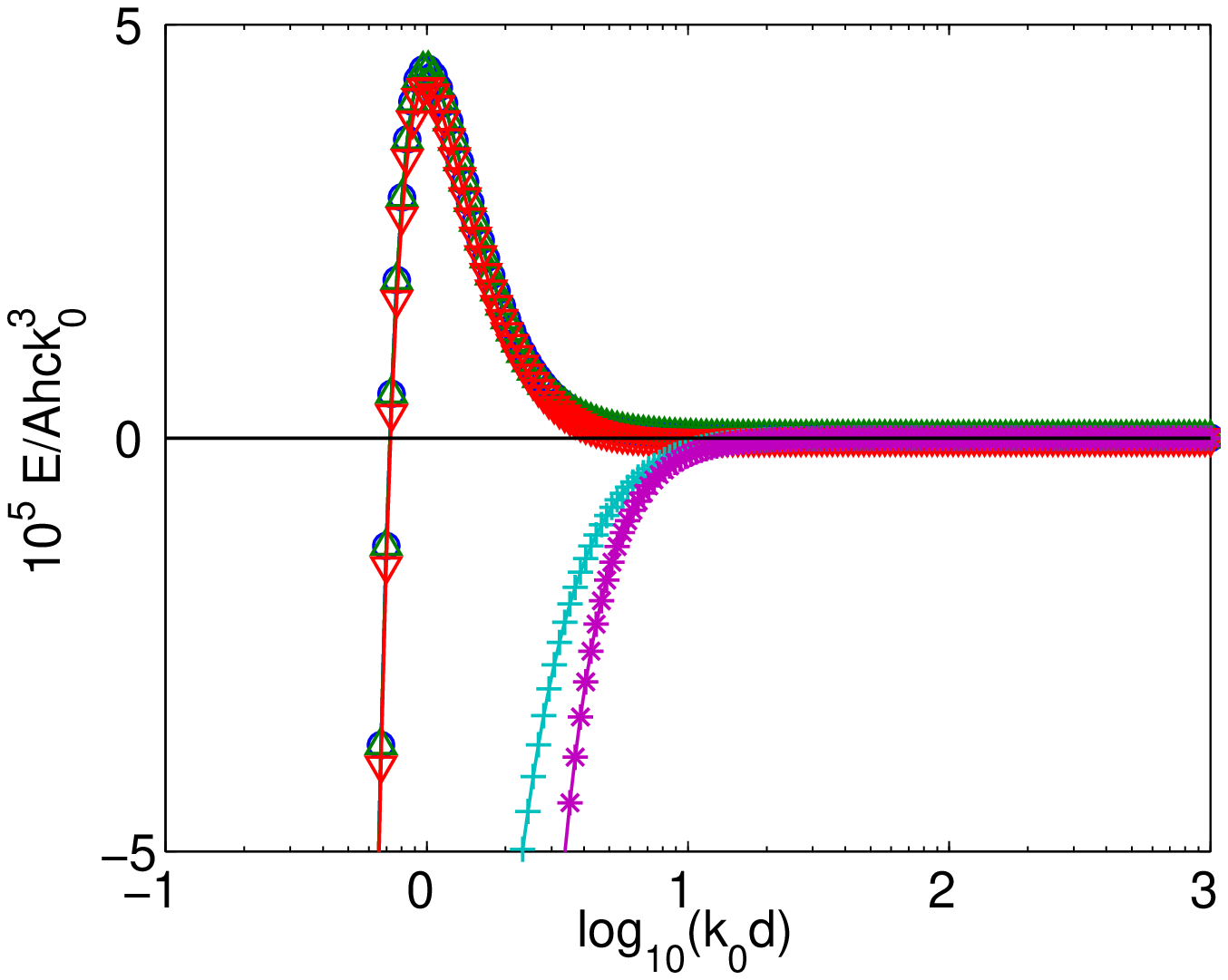}
\put(56,48){\includegraphics[width=0.19\textwidth]{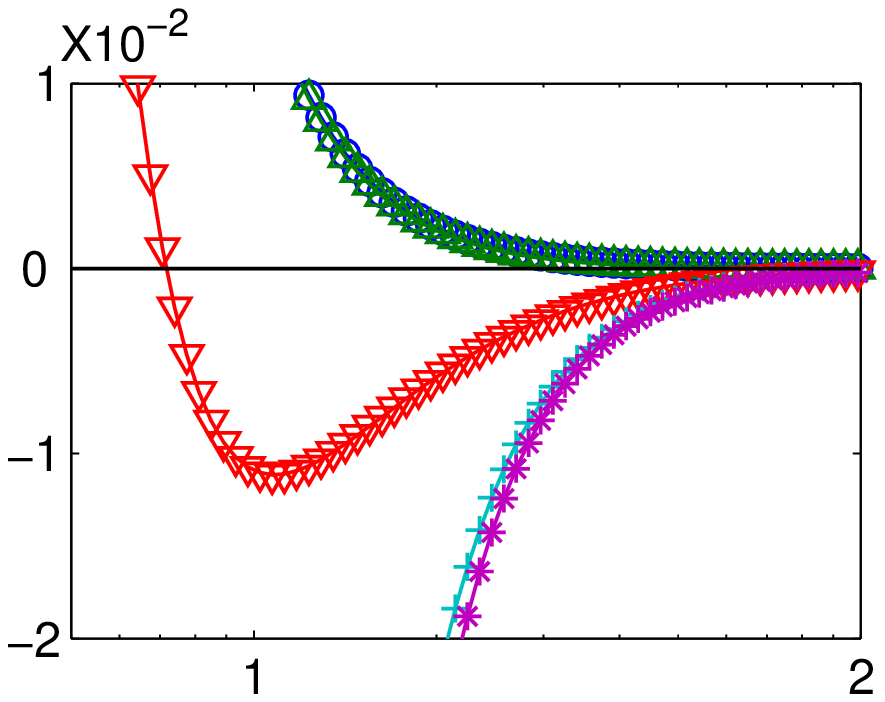}}
\put(56,12){\includegraphics[width=0.19\textwidth]{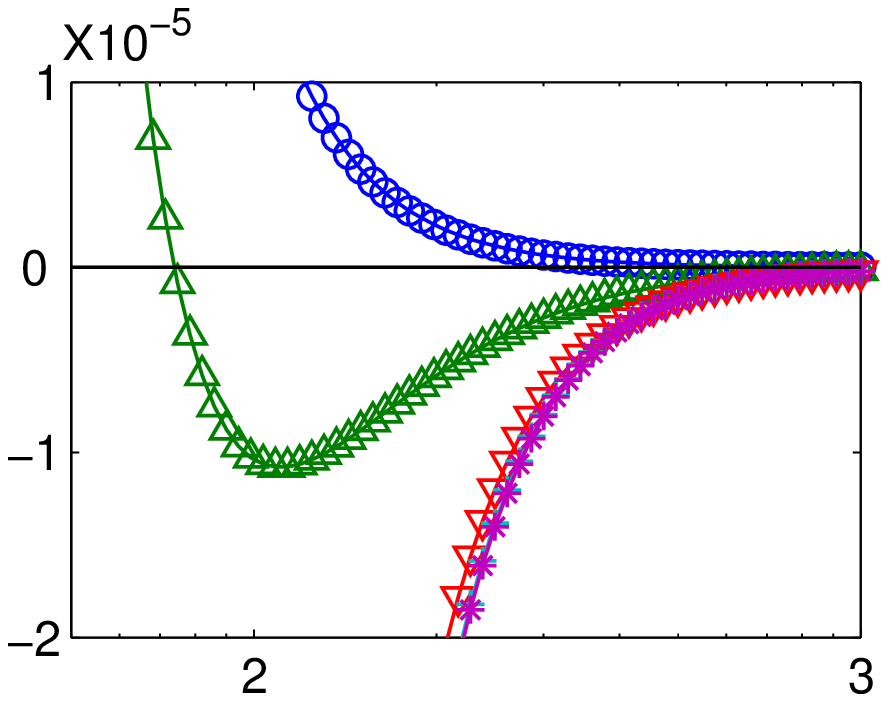}}
\put(16,15){(c)}
\end{overpic}
\hspace{6mm}
\begin{overpic}[width=0.45\textwidth]{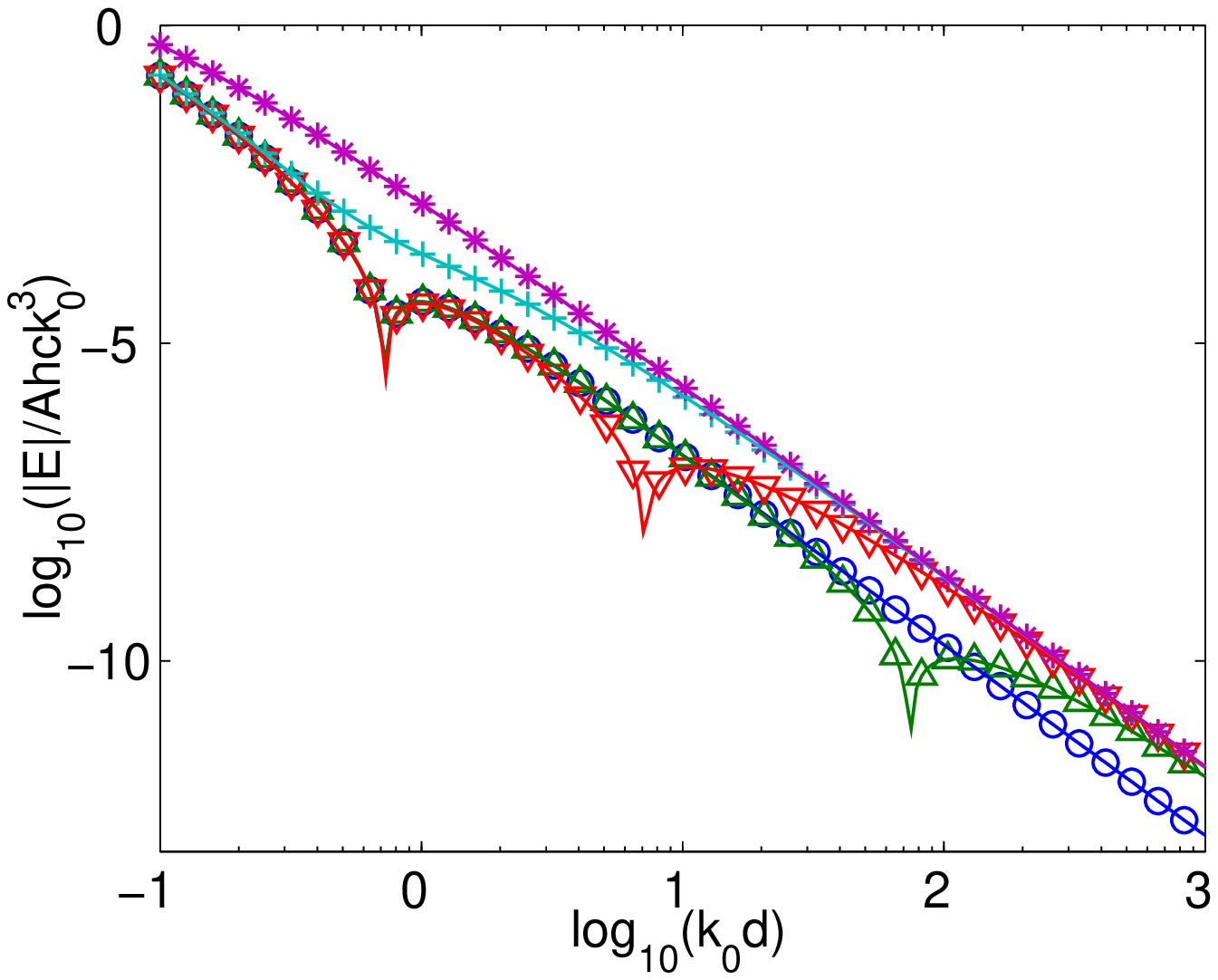}
\put(54,58){\includegraphics[width=0.19\textwidth]{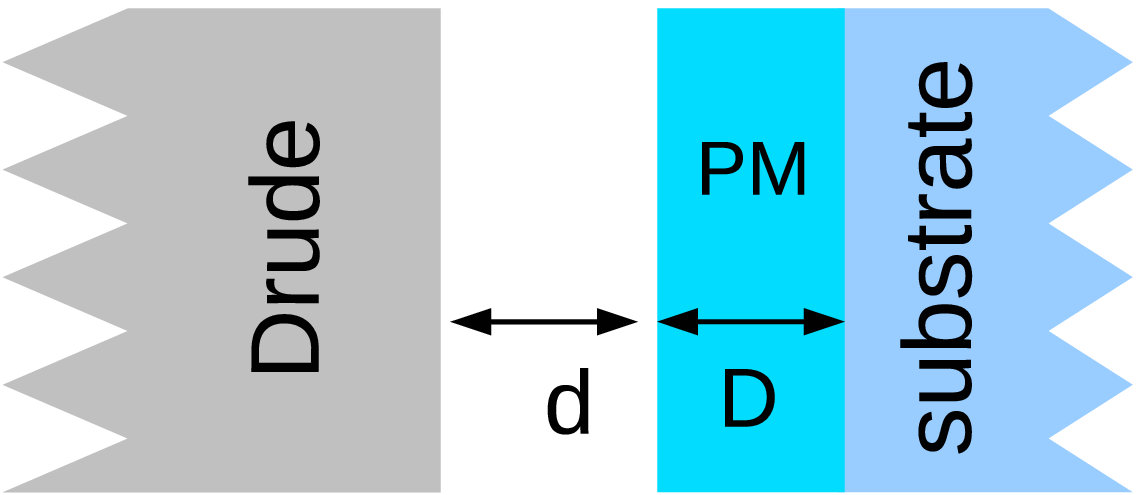}}
\put(16,15){(d)}
\end{overpic}
 \caption{(Color online) Casimir interaction energy per unit area $E(d)/A$ (in unit of $hck_0^3$) versus $k_0d$ between a semi-infinite Drude metal plate and a finite magnetic slab without substrate ((a) and (b)) and with Drude metal substrate ((c) and (d)). The insets in (b) and (d) show the schematic figures depicting the arrangements of the slabs/substrates. The curves correspond to different values of the thickness $D$ of the slab. The linear-log plot (left, (a) and (c) ) and the log-log plot (right, (b) and (d)). The insets in (c) magnify the regions around $k_0d\simeq 10$ and $k_0d\simeq 10^2$.}
 \label{PMenergy}
\end{figure*}

There are claims, e.g., Ref. [27], that when metamaterials are made of ordinary materials
with negligible intrinsic magnetic response, repulsion is impossible at
large distances, but this does not deny the possibility that a paramagnetic slab and and a dielectric slab repel
each other. Yannopapas and his collaborator recently resorted to the magnetic response of paramagnetic composites and obtained theoretically repulsive Casimir force in the micrometer scale.\cite{Yannopapas} Therefore, by employing a proper magnetic response, it is still possible to get a repulsive force. Here we characterize the electric and magnetic response as
\begin{subequations}\label{Drude}
\begin{align}
\epsilon(i\xi)&=1+\frac{\Omega_{\epsilon}\omega_{\epsilon}^2}{\xi^2+\omega_{\epsilon}^2+\gamma_{\epsilon}\xi},\\
\mu(i\xi)&=1+\frac{\Omega_{\mu}\omega_{\mu}^2}{\xi^2+\omega_{\mu}^2+\gamma_{\mu}\xi},
\end{align}
\end{subequations}
where $\Omega_{\epsilon}$ and $\Omega_{\mu}$ denote the strengths of the electric permitivity and magnetic permeability, $\omega_{\epsilon}$ and $\omega_{\mu}$ are the electric and magnetic resonance frequencies, $\gamma_\epsilon$ and $\gamma_\mu$ are the collision frequencies.  However, notice that a $\xi$ dependence of $\mu$ as in Eq.\,(7b) is questionable, since in the existing calculations one obtains that the constant $\Omega_\mu$ is actually replaced by $\Omega_\mu\xi^2$, and the 1 by $1-\Omega_\mu\omega_\mu^2$; the latter because $\mu(+\infty)\to1$. The magnetic response shown in (7b) is assumed to come from the parallel alignment of very small ferromagnetic nanoparticles in an applied magnetic field; therefore, the magnetic resonance frequency is expected to be lower than the electric one. We choose the following parameters: $\omega_{\epsilon}=10\omega_0$, $\omega_{\mu}=\omega_0$, and $\gamma_{\epsilon}=\gamma_{\mu}=0.05\omega_0$, where $\omega_0$ is the normalized frequency. In order to get a repulsive force, the inequality $\Omega_{\mu}>\Omega_{\epsilon}$ should be satisfied. Here, we choose $\Omega_{\mu}=2$ and $\Omega_{\epsilon}=1$ as an example. In the following, we calculate the Casimir force between a semi-infinite Drude metal plate, characterized by $\epsilon(i\xi)=1+\omega_{pl}^2/(\xi^2+\xi \gamma_{pl})$ with $\omega_{pl}=100\omega_0$ and $\gamma_{pl}=0.05\omega_0$, and a magnetic slab of finite thickness $d_2\equiv D$.  Two cases are studied in this paper: (1) with no substrate as shown by the inset of Fig. 1(b); (2) with semi-infinite Drude metal substate as shown by the inset of Fig. 1(d).

\textit{No substrate.}-- Figures \ref{PMenergy}(a) and \ref{PMenergy}(b) show the Casimir interaction energy per unit area $E/A$ versus $k_0d$ between a semi-infinite Drude metal plate and a finite-thickness magnetic slab with no substrate. Different curves correspond to different values of the thickness $D$ of the slab. $k_0=2\pi/\lambda_0$ and $\lambda_0=2\pi c/\omega_0$. These figures show that the Casimir energies exhibit a similar behavior for the slabs of different thicknesses (from $D=+\infty$ to $D=0.01\lambda_0$). Indeed, all Casimir energy curves exhibit a repulsive character for large distances and an attractive one for small distances. Thus, there is only one energy peak (indicating an unstable equilibrium point) appearing approximately at $k_0d\simeq0.7$ for all thickness $D$; the strength of this peak is diminishing as the thickness becomes smaller (especially for $D<0.1\lambda_0$). Figure \ref{PMenergy}(b) shows that, at large distances, the $d$ dependence of the Casimir energy changes from $1/d^3$ (for infinite thickness, $D=+\infty$) to $1/d^4$ (for $D=0.01\lambda_0$); the $1/d^3$ dependence is typical for semi-infinite slabs. Similar $d$ dependence was also found between ordinary media.\cite{Raabe} At small distances, all the Casimir energy curves for different values of thickness (from $D=+\infty$ to $D=0.1\lambda_0$) in Fig. 1(a)  overlap very well. As shown in Fig. 1(b), notice that for $D=10\lambda_0$, the energy curve overlaps with the $D=+\infty$ below $k_0d=10$. If the thickness $D$ becomes smaller, $D=0.1\lambda_0$, the different energy curves overlap below $k_0d=0.1$.

\textit{Drude metal substrate.}-- Figures \ref{PMenergy}(c) and \ref{PMenergy}(d) show the Casimir interaction energy per unit area $E/A$ versus $k_0d$ between a semi-infinite Drude metal plate and a finite magnetic slab with a semi-infinite Drude metal substrate. Different curves correspond to different values of the thickness $D$ of the slab. These figures show that the behavior of the Casimir interaction energies are different for different thicknesses of the slabs. If the finite magnetic slab is very thin (e.g., $D\leq 0.1\lambda_0$ in our case) , the Casimir force is attractive for any distance. A very interesting behavior appears for large (but finite) thicknesses ($D\geq\lambda_0$ in our case): At very large distances $d$ the interaction is attractive (the interaction energy is negative and decreasing with decreasing $d$). Figure \ref{PMenergy}(d) shows that at large distances the $d$ dependence of Casimir energy is $1/d^3$ and does not change with the thickness of the slab. At some distance $d_s$ ($k_0d_s\simeq10$ for $D=\lambda_0$ and $k_0d_s\simeq10^2$ for $D=10\lambda_0$) the interaction energy reaches a local minimum (indicating a stable equilibrium distance)  and then the curve increases as $d$ decreases, it crosses the axis at some point $d_0$ ($d_0<d_s$), it reaches a maximum at $d=d_u$ (at $d_u$ we have an unstable equilibrium distance); $d_u$ seems to be about the same for all thicknesses $D\geq\lambda_0$. For $d<d_u$ the energy curve decreases with decreasing $d$ (indicating an attractive Casimir force) and  crosses the axis at a distance $d_0^\prime$, which seems to be common for all $D\geq\lambda_0$. The points $d_0$ and $d_0^\prime$ are shown as sharp dips in the log-log plot (Fig. 1(d)). It seems that $d_0$  and $d_s$ tend to infinity as $D\to+\infty$.

The appearance of an equilibrium point at the distance $d=d_s$ is of great importance: first, because it can be tuned by the thickness $D$; second, because its magnitude can be larger than the wavelength and/or the size of the unit cell of the magnetic metamaterial and, consequently, it is in the range of validity of the effective medium approximation on which Eqs. (1)-(7b) are based ($d\geq\lambda_0,D$), e.g., for $\lambda_0=700\,\textrm{nm}$ and $D=7\,\mu\textrm{m}$, $d_s\simeq 1 \mu\textrm{m}$.

\section{Repulsive Casimir Forces with Chiral Slabs}
\begin{figure*}[htb!]
\begin{overpic}[width=0.45\textwidth]{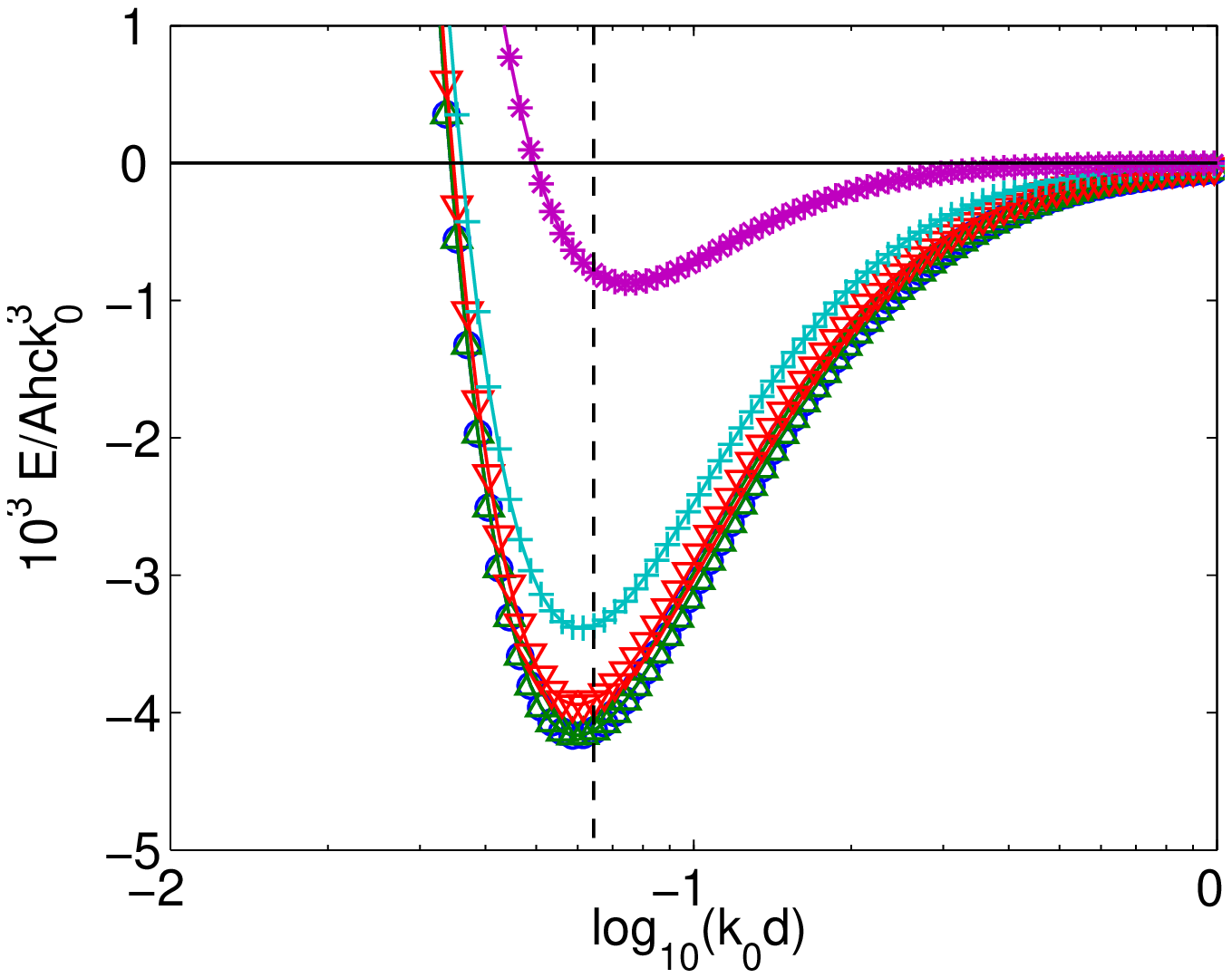}
\put(16,15){(a)}
\end{overpic}
\hspace{6mm}
\begin{overpic}[width=0.45\textwidth]{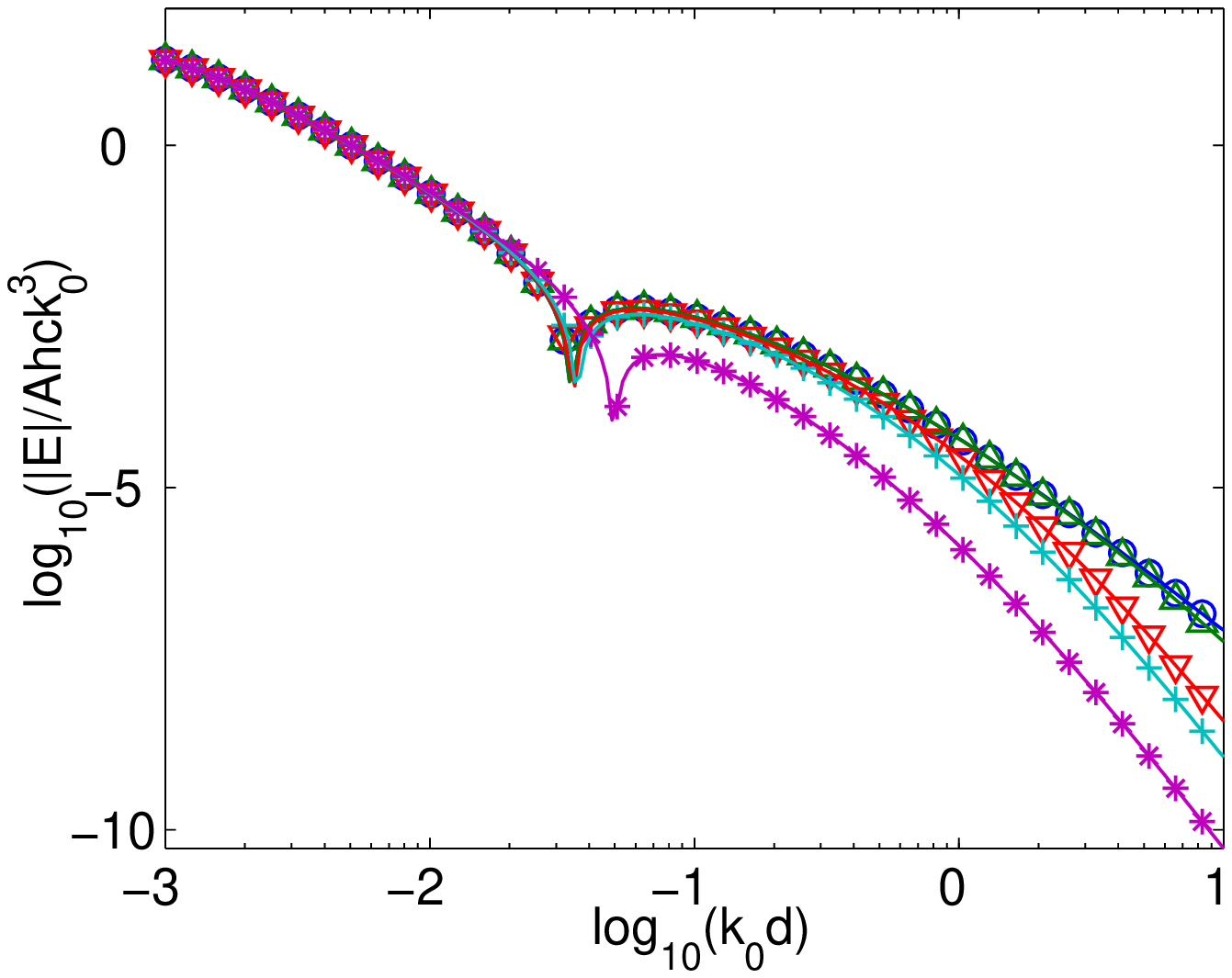}
\put(54,58){\includegraphics[width=0.19\textwidth]{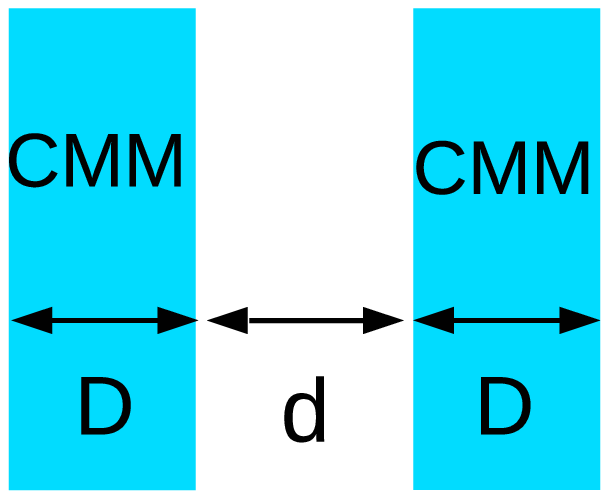}}
\put(16,15){(b)}
\end{overpic}
\begin{overpic}[width=0.45\textwidth]{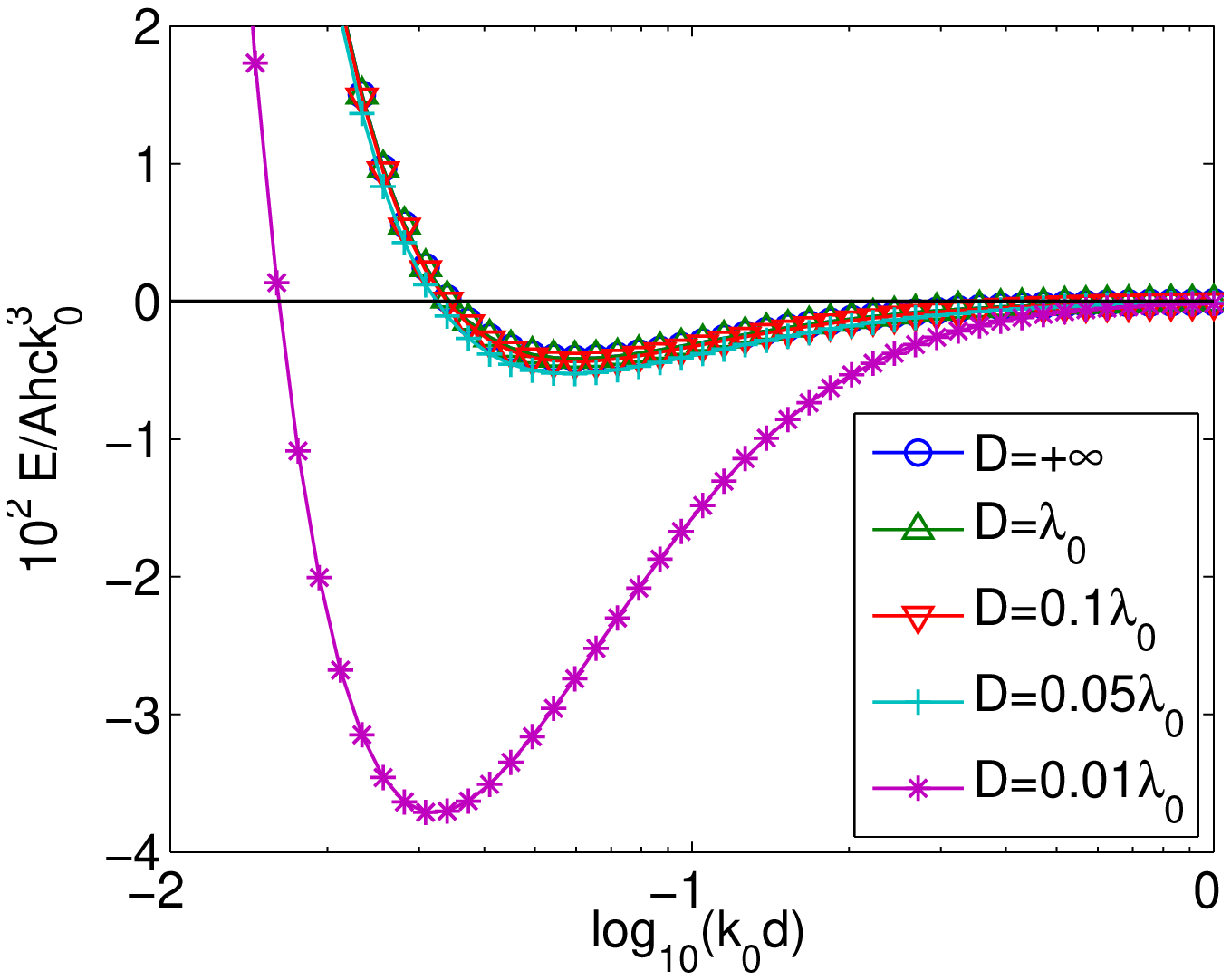}
\put(16,15){(c)}
\end{overpic}
\hspace{6mm}
\begin{overpic}[width=0.45\textwidth]{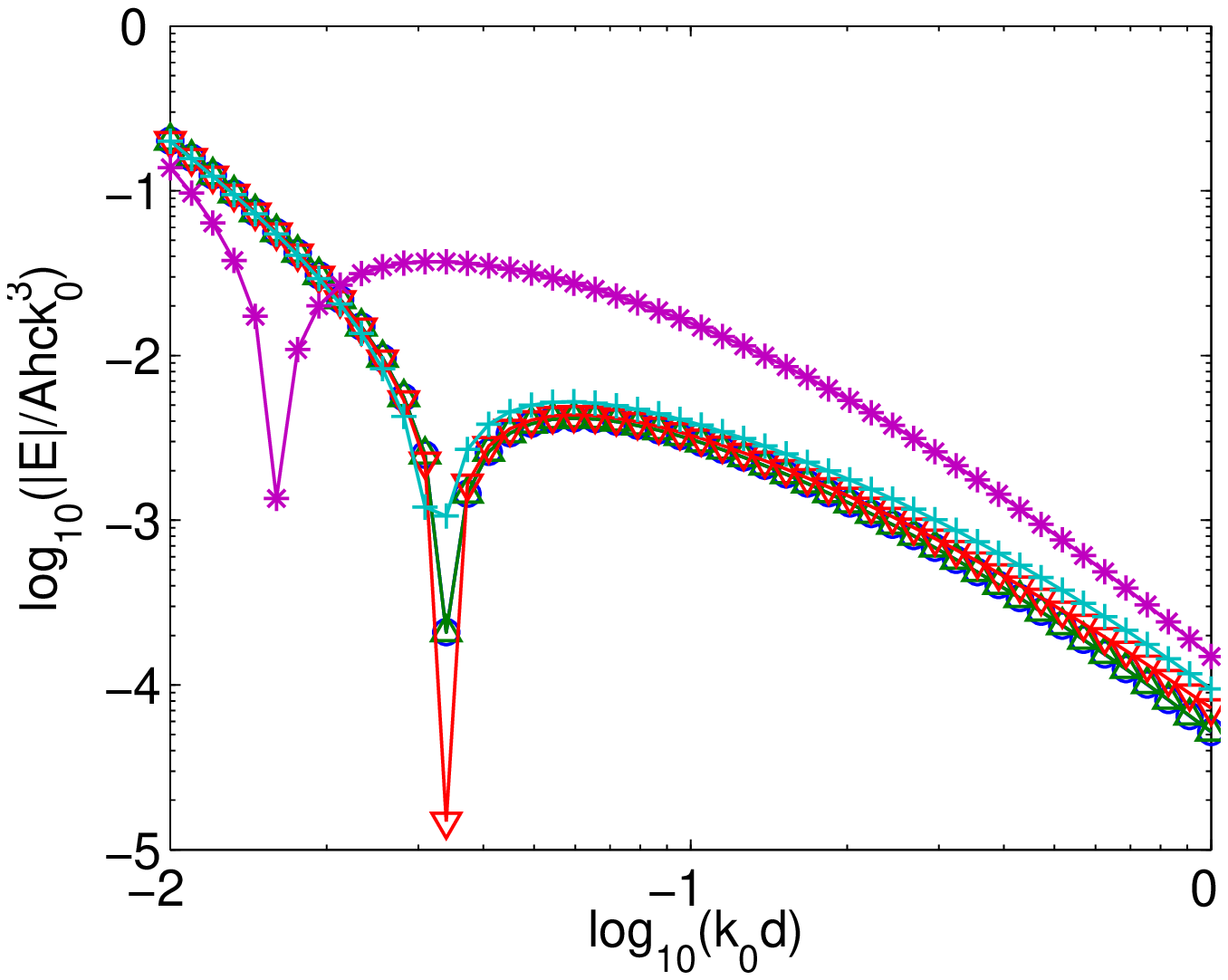}
\put(54,58){\includegraphics[width=0.19\textwidth]{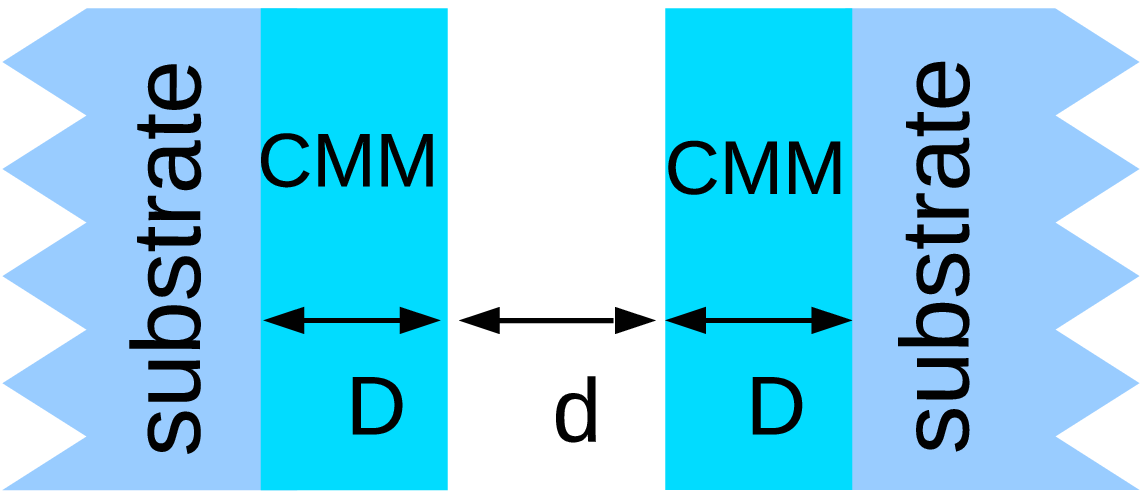}}
\put(16,15){(d)}
\end{overpic}
 \caption{(Color online) Casimir interaction energy per unit area $E(d)/A$ (in unit of $hck_0^3$) versus $k_0d$ between two identical finite-thickness CMM slabs without substrate ((a) and (b)) and with semi-infinite Drude metal substrate ((c) and (d)). The insets in (b) and (d) show the schematic figures depicting the arrangements of the slabs/substrates. The curves correspond to different values of the thickness $D$ of the slab. The linear-log plot (left, (a) and (c) ) and the log-log plot (right, (b) and (d)).}
 \label{CMMenergy}
\end{figure*}
Repulsive Casimir force was also found to be realized by using chiral metamaterials (CMMs) if the chirality is strong enough.\cite{ZhaoPRL,ZhaoPRB} Here, we study the repulsive Casimir force between two finite-thickness chiral metamaterial slabs with or without the Drude metal substrate. The optical parameters of chiral metamaterials are characterized by \cite{ZhaoPRB}
\begin{subequations}\label{CMMparameters}
\begin{align}
\epsilon(i\xi)&=1+\frac{\Omega_{\epsilon}\omega_{\epsilon}^2}{\xi^2+\omega_{\epsilon}^2+\gamma_{\epsilon}\xi},\\
\mu(i\xi)&=1+\Omega_{\mu}-\frac{\Omega_{\mu}\xi^2}{\xi^2+\omega_{\mu}^2+\gamma_{\mu}\xi},\\
\kappa(i\xi)&=\frac{i\Omega_{\kappa}\xi}{\xi^2+\omega_{\kappa}^2+\gamma_{\kappa}\xi},
\end{align}
\end{subequations}
where $\Omega_{\kappa}$ denotes the strength of the chirality resonance, $\omega_{\kappa}$ is the resonance frequency of chiral structure and $\gamma_\kappa$ is the collision frequency. Usually, the electric, magnetic and chirality resonances are at the same frequency, therefore, we set $\omega_\epsilon=\omega_\mu=\omega_\kappa=\omega_0$ and $\gamma_{\epsilon}=\gamma_{\mu}=\gamma_{\kappa}=0.05\omega_0$. In order to get a repulsive force, $\Omega_\kappa$ should be large enough. Here, $\Omega_{\epsilon}=1$, $\Omega_{\mu}=0.001$, and $\Omega_{\kappa}=0.7$, i.e., large enough for repulsive forces to appear. The two slabs are identical with the same parameters and substrate as shown by the insets of Figs. 2(b) and 2(d). Then we still consider two cases: (1) with no substrate as shown by the inset of Fig. 2(b); (2) with semi-infinite Drude metal substate as shown by the inset of Fig. 2(d).

\textit{No substrate.}-- Figures \ref{CMMenergy}(a) and \ref{CMMenergy}(b) show the Casimir interaction energy between two finite chiral slabs without substrate. Different curves correspond to different values of the thickness $D$ of the slabs. We see that no matter how thin (from $+\infty$ to $0.01\lambda_0$) the slabs are, there is only one energy minimum at the whole distance range (from $k_0d=10^{-3}$ to $10$), i.e., in all cases, the Casimir forces have the same behavior: repulsive force at small distances and attractive at large distances as shown in Ref. [10]. All the stable equilibrium points are at around $k_0d=0.07$, where the force changes from attractive to repulsive. As the vertical dashed line at $k_0d=0.06442$ shows, the minimum is at slightly larger distance for small $D$. Notice that the minimum appears at very small distance $d$, which makes the validity of the effective medium theory (EMT) doubtful.\cite{Alex} Figure \ref{CMMenergy}(b) shows similarly that, at large distances, the $d$ dependence of the Casimir energy for finite-thickness slabs is $1/d^5$, as opposed to the  $1/d^3$ between two semi-infinite media. This $d$ dependence was also found theoretically between ordinary slabs.\cite{Raabe}  At small distances, all the Casimir energies for different thicknesses of the finite slabs tend to coincide. However, for such short distances, the EMT is expected to fail and the microstructure effect will dominate the Casimir repulsion effect.\cite{Alex}

\textit{Drude metal substrate.}-- Figures \ref{CMMenergy}(c) and \ref{CMMenergy}(d) show the Casimir interaction energy between two identical finite CMM slabs with the Drude metal substrate. It shows that the behaviors of the Casimir interaction energy are almost the same if the thickness of the slab is larger than $0.05\lambda_0$. The thinner slabs can still give us the repulsive force but at smaller distance, e.g., for $D=0.01\lambda_0$, a repulsive force appears when $k_0d<0.035$. In other words, if we want to demonstrate the Casimir force experimentally, $0.05\lambda_0$ thickness slab is enough to observe all the phenomena, assuming the validity of the EMT, which is doubtful for such short distances, and no microstructure effect. \cite{Alex}

\section{DISCUSSION OF THE VALIDITY OF THE EFFECTIVE MEDIUM APPROXIMATION}
Ref. [29] presents a test of the effective medium approximation (EMA) for chiral metamaterials (as also used in this present paper) against numerical calculations that include the microstructures. A numerical proof was presented \cite{Alex} that the effective homogeneous approximation breaks down when the separation distance between the two plates becomes comparable to the size of the unit cell of the chiral metamaterial making the two plates. On the contrary, we have shown in the present manuscript and in our previous work [10,11], that chirality makes a repulsive contribution to the Casimir force. Our proof \cite{ZhaoPRL,ZhaoPRB} is based on the constitutive equations connecting the Maxwell vectors; these equations are definitely valid in the regime $a \ll t,a \ll d$, where $a$ is the unit cell size of the chiral metamaterial,  $t$ is the thickness of the plates, and $d$ is the separation between the plates. By making $a$  small enough, we can satisfy the double inequality $a \ll d \ll d_0$ , where $d_0$  is a separation distance, such that the Casimir force is appreciable including the chiral repulsive contribution to it. We have shown in the present paper that the combination of plates of finite thickness with appropriate background substantially facilitates the satisfaction of the double inequality. Thus we argue that, because the present manuscript and Ref. [29] consider very different situations, there is no contradiction whatsoever between the two, as explained in detail below.

Previously it has been shown \cite{ZhaoPRL, ZhaoPRB} that two semi-infinite, homogeneous, and isotropic chiral media separated by a finite-thickness vacuum slab will experience a repulsive Casimir force between one another -- or at least a significant reduction of the attractive Casimir force -- at small separations if the chirality of the embedding media becomes large enough. It has been speculated, that such chiral materials could at least, in principle, be implemented by chiral metamaterials in the homogeneous effective medium limit (i.e., where the EMA is valid). The major contribution to the Casimir force comes from frequencies and wave vectors of order of magnitude comparable to the inverse separation of the chiral media; it is in this region at least where the implementation of the chiral metamaterials should allow EMA.

Now, from here chiral Casimir repulsion has been further investigated in at least two directions:

(i) Assume an existing chiral metamaterial structure with a given unit cell size; it has been investigated to what extent a repulsive contribution to the Casimir force can be found (in simulations) for a discreet metamaterial. This is the topic of Ref. [29] where only a minimal repulsive contribution to the Casimir force was found at separation much larger than the unit cell size of the metamaterial -- a regime where also the repulsive contribution in the analytical calculation of homogeneous semi-infinite media would become negligible. Not surprisingly, for separations comparable in order of magnitude to the unit cell size of the metamaterial, it was determined the discrete interactions between the constituents of the metamaterials dominate the force and no chiral repulsion could be observed because the metamaterials do not behave anymore as homogeneous media at the relevant frequencies and wave vectors. Theoretically, this problem could be easily corrected by just making sure the structural length scale, i.e. the assumed unit cell size, is small compared to the separation maintaining the validity of the EMA at the relevant frequencies and wave vectors. Of course, in reality this could be a problem because there are current practical limits to the nano-fabrication of the metamaterial structures (e.g., for repulsion at 1 \textrm{$\mu$}m separation the structural length scale of the metamaterial should be well below 100 nm to ensure homogeneous effective medium behavior). The effect of finite thickness was not studied in [29] -- the media were just chosen thick enough to behave as if they where in fact semi-infinite.

(ii) In this manuscript we follow a very different direction. We keep the assumptions of homogeneous isotropic chiral media and investigate the question how a finite thickness of the semi-infinite chiral media, terminated by air or metal, will affect the sign and magnitude of the Casimir force as well as its scaling with the separation between the media. We consider the homogeneity and isotropy of the chiral materials as given; hence, the implementation by any to-be-designed chiral metamaterial as a technical problem that can be considered independently. We believe this investigation provides valuable information about the physical interplay between Casimir repulsion and attraction brought about by these boundary conditions and is relevant, if an effectively homogeneous metamaterial implementation is fabricated. So, in summary, the present work and Ref. [29] do not contradict each other but shed light on the possibility of a repulsive or reduced magnitude Casimir force from different angles.

We believe the approach taken and results presented here are independent of Ref. [29], not a mere extension of previous work [10,11], and provide new results for finite-thickness effective medium slabs. The discussion of scaling of the Casimir force with separation for the different regimes of thin versus thick finite chiral media slabs, the observation of stable equilibrium points, and the discussion of the effects of different terminations/substrates are unique and important results presented in this manuscript.

Finally, the research reported in this manuscript is in no way "invalidated" by the results reported in Ref. [29]. This previous publication [29] only asserts that once the separation between the chiral media implemented by chiral metamaterials becomes comparable to the structural length scale of the metamaterials, discrete interactions become dominant and the repulsive Casimir force component expected form homogeneous chiral media ceases to exist. Theoretically, the repulsive component to the Casimir force should still exist at any given separation between the chiral media, if only the structural length scale is chosen small enough to ensure validity of the EMA at the relevant frequencies and wave vectors as explained above.

\section{Explanation} Here, we give a physical explanation regarding the Casimir force behaviors shown above: For large distances, the main contribution to the Casimir force comes from the frequencies $\xi<c/d$.\cite{Rosa1} Since $c/d$ is small, the main contribution region comes from low frequencies. For the low frequency waves, the finite thickness of the slab is much less than the wavelengths; therefore, the effective optical parameters of the slab/substrate approach those of the substrate. If the substrate is vacuum, the effective optical parameters of the finite slab approach to those of vacuum; therefore, the Casimir energy of the finite slab with no substrate decreases faster than the traditional Casimir force between two semi-infinite media. Therefore, for the Casimir force between a semi-infinite Drude metal and a finite slab without substrate, the $d$ dependence for large $d$ is $1/d^5$; and for the Casimir force between two identical finite chiral slabs without substrate, the $d$ dependence is $1/d^6$. This behavior of the $d$ dependence is the same as that in the ordinary slabs.\cite{Raabe} If the substrate is Drude metal, the effective optical parameters of the finite slab/substrate for large distances approach to those of Drude metal; therefore, at very large distances, every force approaches to that of the interaction between two semi-infinite Drude metal media, i.e., it is always attractive force at large distance. For short distances, $c/d$ is large. The main contribution region will come from high frequencies (short wavelengths). The influence of the substrate on the finite slab will be small. Therefore, for short distances the slab/substrate system tends to behave as a semi-infinite slab. The interesting behavior appears at intermediate distances for the Drude/magnetic slab/Drude system (see Figs. 1(c) and 1(d)) where the repulsive character of the Drude/slab subsystem competes with the attractive subsystem Drude/Drude. That is why the magnetic slab with a Drude metal substrate can give us the equilibrium point at the intermediate distance. We repeat that for short distances, $c/d$ is large. Hence, the influence of the substrate on the finite slab will be small. Therefore, the finite slab can be considered as a semi-infinite object. As a result, every curve goes to the same value at very small distances. A similar conclusion was given in Ref. [30].  A similar equilibrium point behavior to that shown in the inset of Fig.\,1(c) can also be obtained between two dielectric slabs , $\epsilon_1(i\xi)$ and $\epsilon_2(i\xi)$, sandwiching another liquid $\epsilon_3(i\xi)$ and satisfying the condition of $\epsilon_1(i\xi)<\epsilon_3(i\xi)<\epsilon_2(i\xi)$. For the case of two chiral slabs without substrate, the attractive contribution for large distances (i.e., for low frequencies) is smaller than that of semi-infinite chiral media due to the vacuum substrate, while the repulsive forces at short distances, i.e., for high frequencies almost do not depend on the thickness $D$; therefore, it is easier to obtain the repulsive force, when the latter appears at short distances. Thus for the force between two chiral slabs with Drude metal substrate, because the repulsive contribution comes at very short distances, i.e., for high frequencies, the finite-thickness slab does not influence the repulsive Casimir force too much until $D=0.05\lambda_0$. \\

\section{Conclusion}
In this paper, we used the extended Lifshitz theory to study the repulsive Casimir force between a semi-infinite Drude metal and a finite magnetic slab with or without substrate. For no substrate, we found that at the large distances, the $d$ dependence of the force is $1/d^5$; for the Drude metal substrate, an equilibrium point appears at intermediate distances. The thickness of the slab can tune the position of this equilibrium point. We also study the repulsive Casimir force between two identical chiral slabs with and without substrate. For no substrate, we found that the finite slabs repel each other at short distances, while for large distances the $d$ dependence of the attractive force is $1/d^6$. For the Drude metal substrate, we found that the finite thickness of the slab $D$ does not influence the repulsive force at short distances too much until $D=0.05\lambda_0$. These results are very useful to the experimentalists who are obliged to work with finite slabs.

\section{ACKNOWLEDGMENT}\label{ACKNOWLEDGMENT}
Work at Ames Laboratory was supported by the Department of Energy
(Basic Energy Sciences) under Contract No.~DE-AC02-07CH11358. R.Z.
acknowledges the China Scholarship Council (CSC) for financial support.

\bibliographystyle{apsrev}

\end{document}